\documentclass[preprint,3p,12pt]{elsarticle}

\usepackage{lineno,hyperref}
\modulolinenumbers[5]

\usepackage{hyperref}

\usepackage{epstopdf}

\usepackage{mathrsfs}
\usepackage{amssymb}
\usepackage{amsfonts}
\usepackage{latexsym}
\usepackage{amsmath}
\usepackage{xcolor}
\usepackage{wrapfig}
\usepackage{floatflt}

\usepackage{graphicx}
\def\lb{\label}

\newcommand{\er}[1]{\textrm{(\ref{#1})}}

\newtheorem{theorem}{\bf Theorem}[section]
\newtheorem{lemma}[theorem]{\bf Lemma}
\def\a{\alpha}  \def\cA{{\mathcal A}}       \def\mA{{\mathscr A}}
\def\b{\beta}   \def\cB{{\mathcal B}}       \def\mB{{\mathscr B}}

\def\d{\delta}  \def\cE{{\mathcal E}}       
\def\D{\Delta}  \def\cF{{\mathcal F}}       
           \def\mG{{\mathscr G}}
          
    \def\cI{{\mathcal I}}

\def\l{\lambda}        \def\mM{{\mathscr M}}
 \def\cN{{\mathcal N}}       \def\mN{{\mathscr N}}
\def\m{\mu}            
\def\n{\nu}

    \def\cT{{\mathcal T}}

\def\O{\Omega}

\def\ve{\varepsilon}           

\def\Z{{\mathbb Z}}    \def\R{{\mathbb R}}   \def\C{{\mathbb C}}    
    \def\N{{\mathbb N}}   

\def\lt{\biggl}                  \def\rt{\biggr}
\def\ol{\overline}               \def\wt{\widetilde}
\def\no{\noindent}

\let\ge\geqslant                 \let\le\leqslant

\def\iy{\infty}
\def\sm{\setminus}               \def\es{\emptyset}
\def\ss{\subset}                 \def\ts{\times}
\def\pa{\partial}                
                 \def\ev{\equiv}
        
\def\el2{\ell^{\,2}}             \def\1{1\!\!1}

\def\const{\mathop{\mathrm{const}}\nolimits}
\def\det{\mathop{\mathrm{det}}\nolimits}
\def\diag{\mathop{\mathrm{diag}}\nolimits}

\def\dist{\mathop{\mathrm{dist}}\nolimits}

\def\BBox{\hspace{1mm}\vrule height6pt width5.5pt depth0pt \hspace{6pt}}

\newtheorem{corollary}[theorem]{\bf Corollary}

\let\ge\geqslant
\let\le\leqslant

\newcommand{\ca}{\begin{cases}}
\newcommand{\ac}{\end{cases}}
\newcommand{\ma}{\begin{pmatrix}}
\newcommand{\am}{\end{pmatrix}}
\def\eq{\begin{equation}}
\def\qe{\end{equation}}
\def\[{\begin{equation}}
\def\]{\end{equation}}

\def\BBox{\hspace{1mm}\vrule height6pt width5.5pt depth0pt \hspace{6pt}}

\begin{document}

\begin{frontmatter}

\title{Matrix representations of multidimensional integral and ergodic operators}
%\title{Algebra of mixed multidimensional integral and ergodic operators, a discrete version}
\date{\today}

\author%[A.~A.~Kutsenko]
{Anton A. Kutsenko}%\corref{cor1}}
%\ead{aak@nxt.ru}

%\address{GeoRessources, UMR 7359, ENSG, Vandoeuvre-l\`es-Nancy, 54518,
%France; email: akucenko@gmail.com}
%\address{Department of Mathematics, Aarhus University, Aarhus, DK-8000,
%Denmark; email: akucenko@gmail.com}
\address{Jacobs University (International University Bremen), 28759 Bremen, Germany; email: akucenko@gmail.com}
\address{Saint-Petersburg State University,
Universitetskaya nab. 7/9, St. Petersburg, 199034, Russia}

\begin{abstract}
We provide a representation of the $C^*$-algebra generated by multidimensional integral operators with piecewise constant kernels and discrete ergodic operators. This representation allows us to find the spectrum and to construct the explicit functional calculus on this algebra. The method can be useful in various applications, since many discrete approximations of integral and differential operators belong to this algebra. Some examples are also presented:
1) we construct an explicit functional calculus for extended
Fredholm integral operators with piecewise constant kernels, 2) we
find a wave function and spectral estimates for 3D discrete Schr\"odinger
equation with planar, guided, local potential defects, and point sources. The accuracy of approximation of continuous multi-kernel integral operators by the operators with piecewise constant kernels is also discussed.
%We consider a discrete approximation of the algebra generated by
%multidimensional integral operators and various ergodic operators.
%The explicit representation of this algebra in terms of simple
%matrix algebras is provided. By using this representation, operator
%functions (inverse operators, square roots of operators, etc) and
%the spectrum admit a straightforward computation. As an application:
%1) we construct an explicit functional calculus for extended
%Fredholm integral operators with piecewise constant kernels, 2) we
%find the wave function and spectral estimates for 3D discrete Schr\"odinger
%equation with planar, guided, local potential defects, and point sources.
\end{abstract}

\begin{keyword}
integral equations, functional calculus, Schr\"odnger operator with defects, operator algebras
\end{keyword}

% 31B10, 44A55  primary
% 35P05, 15A69, 74J15, 35J05  secondary

\end{frontmatter}

%\linenumbers

{\section{Introduction}\lb{sec1}}

Let $\O\ss\R^N$ be some domain. Consider the Hilbert space
$L^2:=L^2(\O\to\C^M,\m)$ of vector-valued functions acting on $\O$
($\m$ is the Lebesgue measure). The goal of the paper is the study
of a discrete analogue of the following algebra of operators acting on $L^2$:
\[\lb{001}
 \mA_{\rm c}={\rm Alg}\lt({\bf A}\cdot,\int\cdot dx_1,...,\int\cdot
 dx_N,\cT\rt).
\]
%Because we will focus on the discrete analogue of $\mA_{\rm c}$,
%the definition \er{001} is presented in a brief form.
In other words, $\mA_{\rm c}$ is generated by: 1) multiplication
operators ${\bf A}\cdot$ (the dot $\cdot$ denotes the place of the operator
argument ${\bf u}=(u_m)_{m=1}^M\in L^2$), where ${\bf A}={\bf
A}(x_1,...,x_N)$ are bounded measurable $M\ts M$ matrix-valued functions with complex entries defined on
$\O$; 2) integral operators $\int\cdot dx_n=\int_{\O\cap I}\cdot dx_n$, where $I$ is the one-dimensional set (line) defined by
\[\lb{sect}
 I=I(x_1,...,x_{n-1},x_{n+1},...,x_N)=\{(x_1,...,x_{n-1},y_n,x_{n+1},...,x_N):\ y_n\in\R\},
\]
so, the operator $\int {\bf u}(x_1,...,x_N) dx_n$ integrates ${\bf u}$ in one coordinate $x_n$ producing the function that is constant along this coordinate;  3) operators
\[\lb{covc}
 \cT {\bf u}({\bf x})=(u_m({\bf T}_m({\bf x})))_{m=1}^M,\ {\bf x}\in\O,\ \ {\rm where}\ \ {\bf T}_m:\O\to\O
\]
are measurable mappings. In particular, the algebra $\mA_{\rm c}$
contains various integral operators of Fredholm type, ergodic
operators (which are based on ${\bf T}_m$), and their combinations.
All such operators have different applications in mathematical
physics, e.g., they describe a propagation of waves and other
phenomena in complex structures with defects
\cite{MS,CNJMM,MC,Kjmaa,KIP,C1}, diffusion \cite{N,S2},
thermodynamic processes \cite{PGG,Kr}, random Schr\"odinger
operators and various operators on discrete graphs
\cite{L1,LSV,KS11},  electromagnetic scattering \cite{VFGG}. Some
general aspects of the connection between integral and ergodic
operators are discussed in \cite{EFHN,KP,HI}. Integral operators and some of their finite-dimensional approximations are discussed in \cite{Fredholm,Gbook,AGK}. There are also useful purely algebraic approaches for various integro-differential modules over $C^{\iy}(\R)$ developed in, e.g., \cite{Ro1}, \cite{Bu1}, \cite{GRR}. 

The main problems for
operators from $\mA_{\rm c}$ are to find the spectrum, to find the inverse
operators, square roots, or, more generally, to construct the
functional calculus on this algebra. The difficulty is that
$\mA_{\rm c}$ is very complex. 
%Even there is no explicit representations for operators from $\mA_{\rm c}$. 
We try to find some discrete analogue
$\mA$ of $\mA_{\rm c}$ for which the functional calculus can be
constructed explicitly. One of the most important requirements to
$\mA$ is to be finite dimensional. Because in this case $\mA$ can be
expressed in terms of matrix algebras for which the functional
calculus is well known. If $\mA$ is finite dimensional then, due to
the Stone-Weierstrass theorem, all the matrix-valued functions ${\bf A}$
should be piecewise constant, otherwise the subalgebra generated by ${\bf A}$ has an infinite dimension.
%On the other hand, a wide class of
%${\bf A}$ can be approximated by piece-wise constant functions.
%All
This tells us how the operator algebra $\mA$ should be arranged. It
is natural to suppose that $\O$ is a union of a finite number of
shifted copies of a cube $H=[0,h)^N$ ($h>0$)
\[\lb{002}
 \O=\bigcup_{i=1}^S\O_i,\ \ \O_i={\bf a}_i+H,
\]
where ${\bf a}_i\in\R^N$ are some vertices such that $\O_i$ are disjoint. Recall that $L^2:= L^2(\O\to\C^M,\m)=L^2(\O)^M$ is the Hilbert space of vector-valued functions acting on $\O$, square-integrable with respect to the Lebesgue measure $\m$. For any ${\bf A}\in\C^{M\ts M}$,
$\a\ss\{1,...,N\}$ and $i,j\in\{1,...,S\}$ introduce the following
elementary operators $\cE_{ij}^{\a}[{\bf A}]:L^2\to L^2$ defined by
%\[\lb{003}
% \cE_{ij}^{\a}[{\bf A}]{\bf u}({\bf x}+{\bf a}_i)=\ca h^{-|\a|}{\bf A}\int_{[0,h)^{|\a|}}{\bf u}({\bf x}+{\bf a}_j)d{\bf x}_{\a},& {\bf x}\in H,
% \\ 0,& otherwise,\ac
%\]
\[\lb{003}
 \cE_{ij}^{\a}[{\bf A}]{\bf u}({\bf x})=\ca h^{-|\a|}{\bf A}\int\limits_{\O_i}{\bf u}({\bf x}-{\bf a}_i+{\bf a}_j)d{\bf x}_{\a},& {\bf x}\in \O_i,
 \\ 0,& {\bf x}\in\O\sm\O_i,\ac
\]
where $d{\bf x}_{\a}=\prod_{n\in\a}dx_n$ and the number of elements
in $\a$ is denoted by $|\a|$. Note that $\a$ can be the empty set
$\es$, in this case there is no $\int$ in \er{003}. Also note that $\cE_{ij}^{\a}[{\bf A}]{\bf u}({\bf x})$ is constant along $x_n$ in $\O_i$ for $n\in\a$. In fact, the operator $\cE_{ij}^{\a}[{\bf A}]$ translates the values of ${\bf u}({\bf x})$ from the cube $\O_j$ to the cube $\O_i$, then, it takes the average in $\O_i$ along $x_n$, $n\in\a$ and multiplies the average by ${\bf A}$, and, finally, it puts zero values inside other cubes $\O_r$, $r\ne i$. The examples of action of different operators $\cE$ is demonstrated in Fig. \ref{fig0}. 

\begin{figure}[h]
    \center{\includegraphics[width=0.99\linewidth]{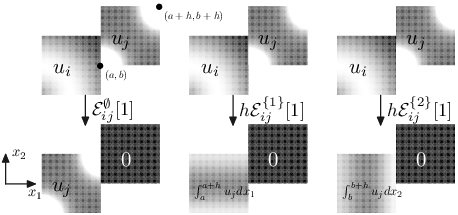}}
    \caption{Schematic view of three different operators $\cE$ in 2D case. While $u({\bf x})$ and $\cE u({\bf x})$ are defined for all ${\bf x}\in\O=\cup\O_r$, the action of $\cE$ is shown for ${\bf x}\in\O_i\cup\O_j$, for simplicity, because $\cE u({\bf x})=0$, ${\bf x}\not\in\O_i$. Here, $u_r$ means $u|_{\O_r}$, $r=i,j$.}\lb{fig0}
\end{figure}

The operators
$\cE_{ij}^{\a}[{\bf A}]$  provide the interaction between the
components of ${\bf u}$ in the various micro-domains $\O_i$.
Roughly speaking, if $h>0$ is sufficiently small then $\O$ can be
approximately represented by \er{002} and the discrete analogue of
$\mA_{\rm c}$ can be chosen as
\[\lb{004}
 \mA={\rm Alg}(\cE_{ij}^{\a}[{\bf A}]).
\]
In other words, $\mA$ is generated by $\cE_{ij}^{\a}[{\bf A}]$ for
all ${\bf A}\in\C^{M\ts M}$, $\a\ss\{1,...,N\}$ and
$i,j\in\{1,...,S\}$.
%(it is sufficient to take only one-element sets
%$\a$ in \er{004}).
Let us discuss why $\mA$ is a discrete analogue of
$\mA_{\rm c}$ defined by \er{001}. The discrete analogues of the multiplication operators ${\bf A}\cdot$ are the multiplication operators with piecewise constant functions ${\bf A}={\bf A}({\bf x})$, i.e.
\[\lb{mod}
 {\bf A}({\bf x})={\bf A}_i=\const,\ \ {\bf x}\in\O_i,\ \ i=1,...,S.
\]
The corresponding operators are expressed in terms of $\cE$ as follows:
\[\lb{mod1}
 {\bf A}\cdot=\sum_{i=1}^S\cE_{ii}^{\es}[{\bf A}_i].
\]
The integral operators, see \er{001} and \er{sect}, are also expressed in terms of $\cE$:
\[\lb{int}
 \int\cdot dx_n=\sum_{i=1}^S\sum_{j\in\b_i}h\cE_{ij}^{\{n\}}[{\bf I}],
\]
where ${\bf I}$ is the identity matrix,
\[\lb{int1}
 \b_i=\{j:\ {\bf a}_j(m)={\bf a}_i(m),\ \forall m\ne n\}
\]
and ${\bf a}(m)$ is the $m$-th entry of the vector ${\bf a}$. The discrete analogues of change-of-variables operators ${\bf T}_m:\O\to\O$ (see \er{covc})  should be based on discrete mappings $p_m:\{1,...,S\}\to\{1,...,S\}$, since $\O=\cup_{i=1}^S\O_i$. Taking arbitrary $p_m$ we construct
\[\lb{cov}
 {\bf T}_m{\bf x}={\bf x}-{\bf a}_i+{\bf a}_{p_m(i)},\ \ {\bf x}\in\O_i,\ \ i=1,...,S.
\]
Then the discrete analogues of $\cT$ \er{covc} are expressed in terms of $\cE$ as follows:
\[\lb{cov1}
 \cT=\sum_{m=1}^M\sum_{i=1}^S\cE_{i,p_m(i)}^{\es}[{\bf I}_m],\ \ {\rm where}\ \ {\bf I}_m=(\d_{im}\d_{jm})_{i,j=1}^{M}\in\C^{M\ts M}
\]
and $\d$ is the Kronecker delta.

The above arguments show that $\mA$ can be considered as the discrete approximation of $\mA_c$.
As it is shown in Theorem \ref{T2} and Example 2 below, if $h\to0$ then
the approximation becomes better and better.
Moreover, $\mA$ is a closed $C^*$-subalgebra of $\mA_{\rm c}$ and, hence,
deserves its own study. {The properties of $\mA$ are enough for most practical applications.}

Denoting by $^*$ the Hermitian conjugation, one can easily check the fundamental relations
\[\lb{005}
 \cE_{ij}^{\a}[{\bf A}]^*=\cE_{ji}^{\a}[{\bf A}^*],\ \ \cE_{ij}^{\a}[{\bf A}]+\cE_{ij}^{\a}[{\bf B}]=\cE_{ij}^{\a}[{\bf A}+{\bf B}],\ \
 \cE_{ij}^{\a}[{\bf A}]\cE_{kl}^{\b}[{\bf B}]=\d_{jk}\cE_{il}^{\a\cup\b}[{\bf A}{\bf
 B}].
\]
Hence, $\cE_{ij}^{\a}[{\bf A}]$
are basis elements and any operator $\cA\in\mA$ has the form
\[\lb{006}
 \cA=\sum_{\a\ss\{1,...,N\}}\sum_{i,j=1}^S\cE_{ij}^{\a}[{\bf
 A}_{ij}^{\a}],
\]
where ${\bf A}_{ij}^{\a}$ are some $M\ts M$ matrices. In practice,
form \er{006} is available after taking an approximation of the
initial operator $\cA_{\rm c}\in\mA_{\rm c}$. The question is
how to find explicitly the spectrum of $\cA$, inverse $\cA^{-1}$,
square root $\sqrt{\cA}$, etc. As mentioned above, if we provide a
representation of $\mA$ in terms of simple matrix algebras, then the
answers on all these questions become explicit. We denote
the simple matrix algebras as $\C^{n\ts n}$, $n\ge1$. Introduce the
following matrices
\[\lb{007}
 {\bf A}_{\a}=({\bf A}_{ij}^{\a})_{i,j=1}^S\in\C^{MS\ts MS},\ \ {\bf
 B}_{\a}=\sum_{\b\ss\a}{\bf A}_{\b}
\]
and the following mapping
\[\lb{008}
 \pi:\mA\to(\C^{MS\ts MS})^{2^N},\ \ \pi(\cA)=({\bf
 B}_{\a})_{\a\ss\{1,...,N\}},
\]
where $\cA$ is given by \er{006}. The next theorem is our main result.
\begin{theorem}\lb{T1}
The mapping $\pi$ is the $C^*$-isomorphism between $C^*$-algebras $\mA$ and $(\C^{MS\ts MS})^{2^N}$. The inverse mapping has the form
\[\lb{009}
 \pi^{-1}(({\bf B}_{\a})_{\a\ss\{1,...,N\}})=\sum_{\a\ss\{1,...,N\}}\sum_{i,j=1}^S\cE_{ij}^{\a}[{\bf
 A}_{ij}^{\a}],
\]
where ${\bf A}_{ij}^{\a}\in\C^{M\ts M}$ are the blocks of the matrix ${\bf A}_{\a}=({\bf A}_{ij}^{\a})_{i,j=1}^S$ given by
\[\lb{010}
 {\bf A}_{\a}=
 \sum_{\b\ss\a}(-1)^{|\a\sm\b|}{\bf B}_{\b}.
\]
\end{theorem}
Note that while the most of operators from $\mA$ are infinite-dimensional and even non-compact, the algebra $\mA$ has a finite dimension. We immediately obtain the following
\begin{corollary}\lb{C1}
i) The operator $\cA$ is invertible if and only if all the matrices
${\bf B}_{\a}$ are invertible. In this case, $\cA^{-1}$ can be
computed explicitly
\[\lb{011}
 \cA^{-1}=\pi^{-1}(({\bf B}_{\a}^{-1})_{\a\ss\{1,...,N\}}).
\]

ii) Generalizing i) we can take rational functions $f$ and write
\[\lb{012}
 f(\cA)=\pi^{-1}((f({\bf B}_{\a}))_{\a\ss\{1,...,N\}}).
\]
The extension to algebraic and transcendent functions $f$ is also
obvious.

iii) The trace and the determinant of $\cA$ can be defined as
\[\lb{012a}
 \det(\cA)=\prod_{\a\ss\{1,...,N\}}\det{\bf B}_{\a},\ \ \ {\rm tr}(\cA)=\sum_{\a\ss\{1,...,N\}}{\rm tr}{\bf
 B}_{\a},
\]
they satisfy the usual properties
\[\lb{012b}
 \det(\cA\cB)=\det(\cA)\det(\cB),\ \ \ {\rm tr}(a\cA+b\cB)=a{\rm
 tr}(\cA)+b{\rm tr}(\cB),
\]
\[\lb{012c}
 \det(e^{\cA})=e^{{\rm tr}(\cA)},\ \ \ {\rm sp}(\cA)=\{\l:\
 \det(\cA-\l\cI)=0\},
\]
where $\cI=\pi^{-1}(({\bf I})_{\a\ss\{1,...,N\}})$ is the identity
operator, ${\bf I}$ is $MS\ts MS$ identity matrix, and $a,b\in\C$, $\cA,\cB\in\mA$.

iv) Since $\pi$ is the $C^*$-isomorphism, the operator norm of $\cA\in\mA$ can be computed explicitly
\[\lb{012d}
 \|\cA\|_{L^2\to L^2}=\max\{\l:\ \l\ are\ s-values\ of\ {\bf B}_{\a}\},
\]
where, recall that $\pi(\cA)=({\bf B}_{\a})_{\a\ss\{1,...,N\}}$.
\end{corollary}

\no{\bf Remark.} The algebra 
$$
 \mB={\rm Alg}\lt({\bf A}\cdot,\int_0^1\cdot dx_1,...,\int_0^1\cdot dx_n\rt)
$$
with piecewise constant functions ${\bf A}$ acting on $\O=[0,1)^N$ is considered in \cite{AK11}. The difference between $\mA$ and $\mB$ is the presence of more general class of sets $\O$ and the addition of change-of-variables operators $\cT$. In the case $\O=[0,1)^N$ with  the uniform partition of each of the segments $[0,1)$ on $p$ intervals (i.e. $S=p^N$), $\mB$ becomes a $C^*$-subalgebra of $\mA$:
$$
 \mB\cong\prod_{n=0}^N(\C^{Mp^n\ts Mp^n})^{{\binom {N} {n}}p^{N-n}}\ss(\C^{Mp^N\ts Mp^N})^{2^N}\cong\mA,
$$
where ${\binom {N} {n}}$ are binomial coefficients. It is interesting to note that the generalization of the structure of $\mA$ simplifies the proof of main results.

Let us discuss a norm of approximation of continuous operators from $\mA_{\rm c}$ by discrete operators from $\mA$. For simplicity, consider the case $\O=[0,1)^N$ and $M=1$. The generalization to $M>1$ is similar. Consider a multi-kernel operator $\cA_c:L^2(\O)\to L^2(\O)$ of the form, common in applications,
\[\lb{o13}
 \cA_c u({\bf k})=\sum_{\a\ss\{1,...,N\}}\int_{[0,1)^{|\a|}}A_{\a}({\bf k},{\bf x}_{\a})u({\bf k}_{\ol{\a}}+{\bf x}_{\a})d{\bf x}_{\a},\ \ {\bf k}\in\O,\ \  u\in L^2(\O).
\] 
Here, we also use the notation
$$
 {\bf x}_{\a}=(\wt x_n),\ \ \wt x_n=\ca x_n, & n\in\a,\\ 0,& n\not\in\a. \ac
$$
Note that ${\bf x}={\bf x}_{\a}+{\bf x}_{\ol{\a}}$, where $\ol{\a}=\{1,...,N\}\sm\a$ is the complement to the set $\a$. Consider the uniform partition of $\O$ onto $p^N$ identical cubes
\[\lb{o14}
 \O=\bigcup_{i=1}^{p^N}\O_i,\ \ \O_i={\bf a}_i+[0,1/p)^N,
\] 
where
\[\lb{o15}
 {\bf a}_i=\frac1{p^N}(b_j)_{j=1}^N,\ \ i-1=\sum_{j=1}^Np^{j-1}b_j,\ \ b_j\in\{0,...,p-1\}
\]
is the representation of $i-1$ in the base $p$ numeral system.
Let us take the approximation of $\cA_c$ by
\[\lb{o16}
 \cA=\sum_{\a\ss\{1,...,N\}}\sum_{i=1}^{p^N}\sum_{j=1}^{p^N}\frac1{p^{|\a|}}\cE^{\a}_{ij}[A^{\a}_{ij}],
\]
where 
\[\lb{o17}
 A^{\a}_{ij}=\ca A_{\a}({\bf a}_i+\pmb{\ve},({\bf a}_j+\pmb{\ve})_{\a}), & ({\bf a}_i)_{\ol{\a}}=({\bf a}_j)_{\ol{\a}},\\
 0,& otherwise,\ac\ \ {\rm and}\ \  \pmb{\ve}=(1/(2p))_{r=1}^N.
\]
The next proposition shows us that $\cA\to\cA_c$ in the operator norm.
\begin{theorem}\lb{T2}
Suppose that $A_{\a}({\bf k},{\bf x}_{\a})$ in \er{o13} are real functions with bounded first derivatives. Then
\[\lb{o18}
 \|\cA_c-\cA\|_{L^2\to L^2}\le\frac{N}{2p}\sum_{\a\ss\{1,...,N\}}\|\nabla A_{\a}\|_{L^{\iy}}\to0\ for\ p\to\iy.
\]
\end{theorem}
Along with Theorem \ref{T1} and Corollary \ref{C1}, we can use Theorem \ref{T2} to determine the spectrum and the inverse operator.
\begin{corollary}\lb{C2}
Under the assumptions of Theorem \ref{T2}, we assume also that $\cA_c$ is self-adjoint, i.e. $A_{\a}({\bf k},{\bf x}_{\a})=A_{\a}({\bf k}_{\ol{\a}}+{\bf x}_{\a},{\bf k}_{\a})$. Then $\cA$ is self-adjoint and
\[\lb{o19}
 {\rm sp}(\cA_c)\ss B_{\d}({\rm sp}(\cA)),\ \ {\rm sp}(\cA)\ss B_{\d}({\rm sp}(\cA_c)),
\]  
where $B_{\d}$ means $\d$-neighbourhood and we can set $\d=\frac{N}{2p}\sum_{\a\ss\{1,...,N\}}\|\nabla A_{\a}\|_{L^{\iy}}$. If $0\not\in{\rm sp}(\cA_c)$ then $\cA_c$ is invertible and there is $p$ such that $0\not\in{\rm sp}(\cA)$ and $\|\cA^{-1}\|_{L^2\to L^2}\le\d^{-1}$, and the following estimate is true
\[\lb{o20}
 \|\cA_c^{-1}-\cA^{-1}\|_{L^2\to L^2}\le\frac{\d\|\cA^{-1}\|^2_{L^2\to L^2}}{1-\d\|\cA^{-1}\|_{L^2\to L^2}}.
\] 
Moreover, RHS of \er{o20} tends to 0 for $p\to\iy$.
\end{corollary}

The rest of the paper is organized as follows. Section \ref{Examples} contains
two examples: 1) new formulas for the functions of 1D Fredholm
integral operators with step kernels; 2) the application of the
method for obtaining a solution (with arbitrary precision) and
spectral estimates for 3D discrete Schr\"odinger equation with
planar, guided, and local potential defects. A short proof of the main
result based on the explicit representation of a semigroup algebra
of subsets is given in Section \ref{Proof}. We conclude in Section \ref{Conclusion}.

\section{Examples}\lb{Examples}

{\bf Example 1.}
%An example should be simple to show that the
%approach is general.
Consider the case $N,M=1$, and the
classical Fredholm integral operators (see \cite{Fredholm})
\[\lb{013}
 \cA:L^2([0,1))\to L^2([0,1)),\ \ \cA u(x)=\int_0^1B(x,y)u(y)dy
\]
with $S$-step (piecewise constant) kernels
\[\lb{014}
 B(x,y)=S\sum_{i,j=1}^{S}B_{ij}\chi_i(x)\chi_j(y),\ \ B_{ij}\in\C,\ \
 \chi_i(x)=\ca 1, & x\in[\frac{i-1}S,\frac iS),\\ 0,& otherwise. \ac
\]
Such operators form an algebra isomorphic to $\C^{S\ts S}$ (see, e.g.,
\cite{Gbook}). But, this algebra does not contain the identity
operator ($\cI u=u$). Let us supplement it by adding new operators
\[\lb{015}
 \cA u(x)=A(x)u(x)+\int_0^1B(x,y)u(y)dy,\ \ 
%\]
%\[\lb{016}
 where\ \ A(x)=\sum_{i=1}^SA_i\chi_i(x).
\]
In other words
\[\lb{017}
 A(x)=\pmb{\chi}^{\top}(x){\bf A}\pmb{\chi}(x),\ \ B(x,y)=S\pmb{\chi}^{\top}(x){\bf
 B}\pmb{\chi}(y),\ \ where
\]
\[\lb{018}
 \pmb{\chi}=(\chi_i)_{i=1}^S,\ \ {\bf A}=\diag(A_{i}),\ \ {\bf
 B}=(B_{ij})_{i,j=1}^S.
\]
Taking ${\bf a}_i=(i-1)/S$ and using notations \er{002}-\er{003},
we obtain
\[\lb{019}
 \cA=\sum_{i=1}^S\cE_{ii}^{\es}[A_i]+\sum_{i,j=1}^S\cE_{ij}^{\{1\}}[B_{ij}].
\]
Applying the results of Theorem \ref{T1} and Corollary \ref{C1}
along with \er{015}-\er{018} we obtain that
\[\lb{020}
 \cA^{-1}u(x)=\pmb{\chi}^{\top}(x){\bf A}^{-1}\pmb{\chi}(x)u(x)+S\int_0^1\pmb{\chi}^{\top}(x)(({\bf A}+{\bf
 B})^{-1}-{\bf A}^{-1})\pmb{\chi}(y)u(y)dy,
\]
if and only if ${\bf A}$ and ${\bf A}+{\bf B}$ are invertible (otherwise $\cA$ is non-invertible). The spectrum of $\cA$ consists of all the eigenvalues
of ${\bf A}$ and ${\bf A}+{\bf B}$. In general,
\[\lb{021}
 f(\cA)u(x)=\pmb{\chi}^{\top}(x)f({\bf A})\pmb{\chi}(x)u(x)+S\int_0^1\pmb{\chi}^{\top}(x)(f({\bf A}+{\bf
 B})-f({\bf A}))\pmb{\chi}(y)u(y)dy
\]
for various functions $f$ (e.g.
rational/algebraic/transcendent:$\exp,\sqrt{\ },...$). We have reduced
explicitly the functional calculus on integral operators to the
functional calculus on matrices. 

Let us consider a concrete example with some oscillating non-symmetric integral kernel, say
\[\lb{cexamp1}
 \cA_{\rm c} u(x)=A(x)u(x)+\int_0^1B(x,y)u(y)dy,
\] 
where
\[\lb{cexamp2}
 A(x)=-x\sin(\pi x)-1,\ \ \ B(x,y)=4\sin(6\pi xy)+\exp(x)\sqrt{y+2}.
\]
Taking the step approximation of $A$ and $B$, see \er{014}-\er{018}, with sufficiently large $S$, and using Corollary \ref{C1}, we can estimate the eigenvalue $\l_0$ closest to $0$. It is $\l_0\approx-0.1\ne0$. Hence, the operator $\cA_{\rm c}$ has an inverse. Note that we can not use a Neumann series for $\cA_{\rm c}^{-1}$, since the kernel $B(x,y)$ is not small relative to the first term $A(x)$.  Let us approximately compute $\cA_{\rm c}^{-1}$ and, say, $\cA_{\rm c}^{\frac13}$, where the real branch of cubic root is assumed. To do this, we apply \er{021} to the step approximations of $A$ and $B$. Thus,
\[\lb{cexamp3}
 \cA_{\rm c}^{-1}u(x)=\frac{u(x)}{A(x)}+\int_0^1C(x,y)u(y)dy,\ \ \cA_{\rm c}^{\frac13}u(x)=(A(x))^{\frac13}u(x)+\int_0^1D(x,y)u(y)dy.
\]
%To show that integral kernels $C$ and $D$ exist, we can not use directly Corollary \ref{C2}, since $A$ is non-symmetric. Nevertheless, the
 Computations show a good convergence of step kernels obtained by \er{021}. For example, the relative deviations computed for $S=60$, $300$, and $1500$ are sufficiently small
$$
 \frac{\|C_{60}-C_{300}\|_{L^{\iy}}}{\|C_{300}\|_{L^{\iy}}}=0.1,\ \ \frac{\|C_{300}-C_{1500}\|_{L^{\iy}}}{\|C_{300}\|_{L^{\iy}}}=0.02,
$$
$$ 
   \frac{\|D_{60}-D_{300}\|_{L^{\iy}}}{\|D_{300}\|_{L^{\iy}}}=0.07,\ \ 
 \frac{\|D_{300}-D_{1500}\|_{L^{\iy}}}{\|D_{300}\|_{L^{\iy}}}=0.01,
$$
while
$$
 \|C_{60}\|_{L^{\iy}}\approx\|C_{300}\|_{L^{\iy}}\approx\|C_{1500}\|_{L^{\iy}}\approx30,\ \ \|D_{60}\|_{L^{\iy}}\approx\|D_{300}\|_{L^{\iy}}\approx\|D_{1500}\|_{L^{\iy}}\approx5.
$$
The corresponding integral kernels are plotted in Fig. \ref{kernels}.
\begin{figure}[h]
	\center{\includegraphics[width=1\linewidth]{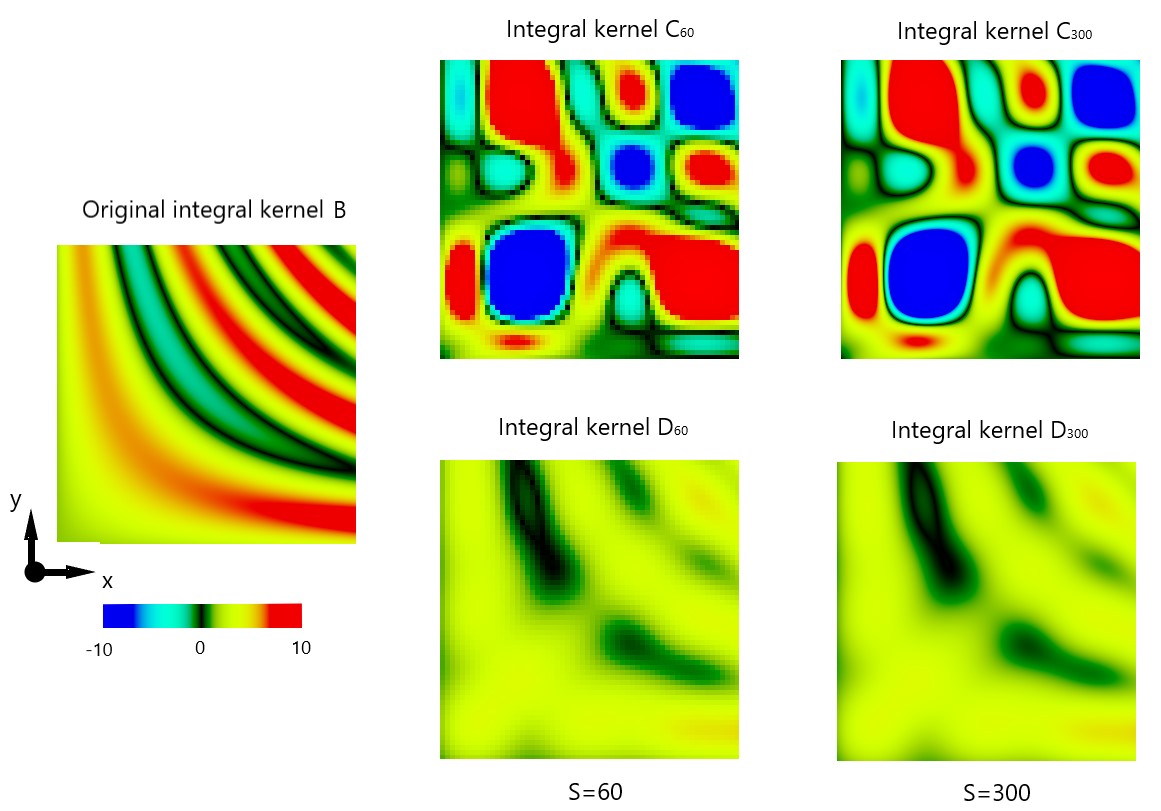}}
	\caption{The original integral kernel $B(x,y)$, see \er{cexamp2}, and the step approximations of integral kernels $C(x,y)$, $D(x,y)$, see \er{cexamp3}, computed by \er{021} for $S=60$ and $S=300$.}\lb{kernels}
\end{figure}

{\bf Example 2.}
\begin{figure}[h]
    \center{\includegraphics[width=0.7\linewidth]{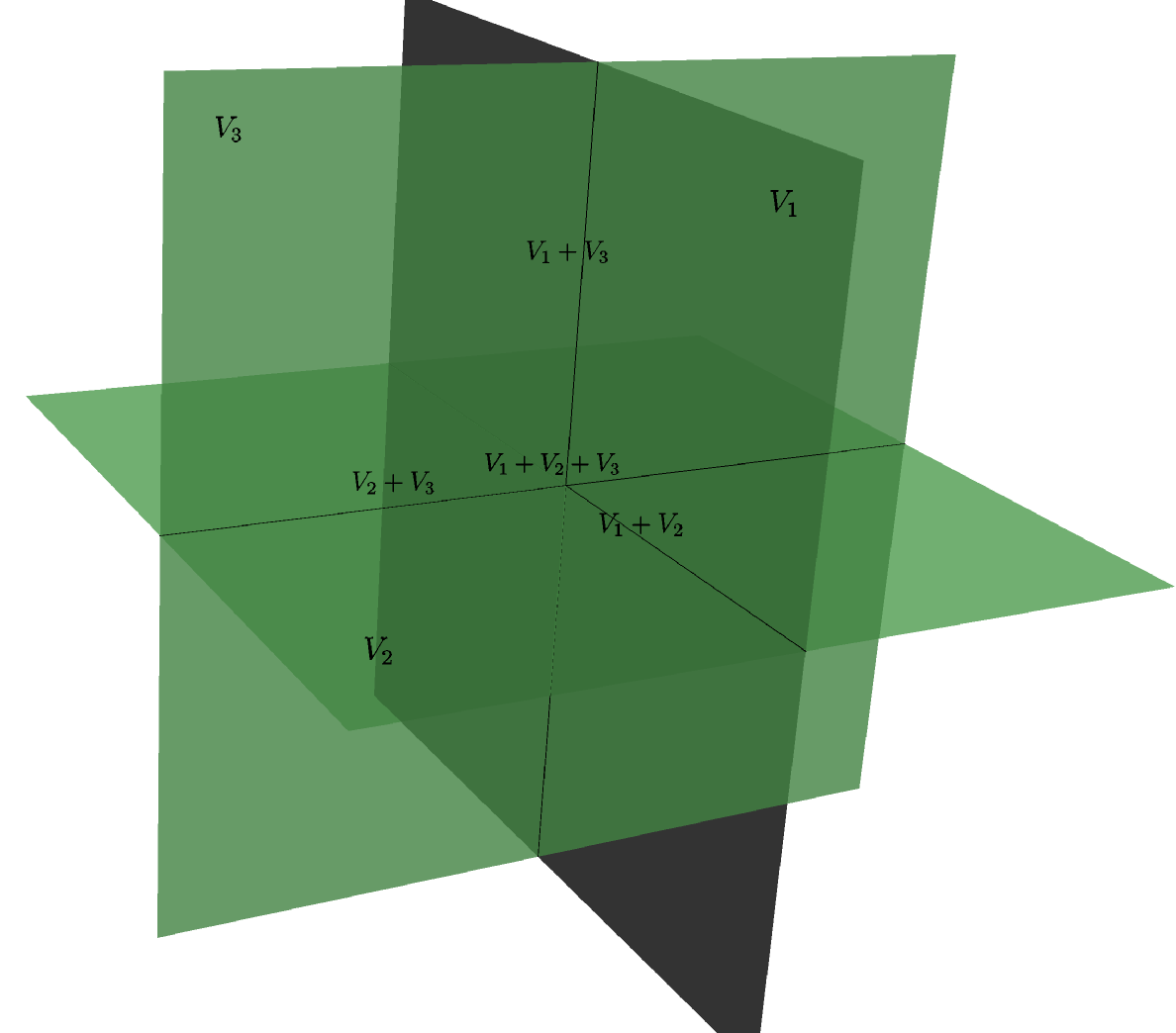}}
    \caption{The lattice $\Z^3$ with planar potential defects $V_i$, line potential
    defects $V_i+V_j$, and point defect $V_1+V_2+V_3$.}\lb{fig1}
\end{figure}
Actually, the method based on the
expansion in $\cE_{nm}^{\alpha}$ works well for more complex mixed
multidimensional integral operators. Let us consider a discrete normalized 3D
time-dependent Schr\"odinger operator with three planar potentials
$V_1$, $V_2$, $V_3$ with their intersections (see Fig. \ref{fig1})
and with attenuated harmonic source term $e^{-(i\l+\ve)t}$
($\l\in\R$ is the frequency, $\ve>0$ is the attenuation factor, and
the amplitude is $1$) located at the origin. Then the corresponding
equation on the wave function $\Psi$ is
\[\lb{e001}
 i\frac{\pa\Psi_{\bf n}}{\pa t}-
 \D_{discr}\Psi_{\bf n}-V_{\bf n}\Psi_{\bf n}=
 \d_{{\bf n}{\bf 0}}e^{-(i\l+\ve)t},\ \ {\bf n}=(n_i)_{i=1}^3\in\Z^3,
\]
where
\[\lb{e002}
 V_{\bf n}=\ca V_i,& n_i=0,\ \prod_{j\ne i} n_j\ne0,\\
 V_i+V_j,&n_i=n_j=0,\ n_k\ne0\ (k\not\in\{i,j\})\\ V_1+V_2+V_3,& {\bf n}=0,\\ 0,& otherwise \ac,\
 \ \ \ \D_{discr}\Psi_{\bf n}=\sum_{\wt{\bf n}}(\Psi_{\bf n}-\Psi_{\wt{\bf n}})
\]
and $\wt{\bf n}$ denotes points adjacent to ${\bf n}$. Our goal is to solve the Schr\"odinger equation and
to find $\Psi_{\bf n}$. Assuming $\Psi_{\bf n}(t)=\hat\Psi_{\bf
n}e^{-(i\l+\ve)t}$ with time-independent $\hat\Psi_{\bf n}$ and
taking Fourier series $\hat\Psi({\bf k})=\sum_{{\bf
n}\in\Z^3}e^{2\pi i{\bf n}^{\top}{\bf k}}\hat\Psi_{\bf n}$, ${\bf
k}=(k_i)_{i=1}^3\in[0,1)^3$ we can rewrite \er{e001} in the integral
form
\begin{multline}\lb{e003}
 A({\bf k})\hat\Psi({\bf
 k})\\-V_1\int_0^1\hat\Psi(x_1,k_2,k_3)dx_1-
 V_2\int_0^1\hat\Psi(k_1,x_2,k_3)dx_2-V_3\int_0^1\hat\Psi(k_1,k_2,x_3)dx_3=1,
\end{multline}
where $%\[\lb{e004}
 A({\bf k})=-i\ve+\l-4\sum_{i=1}^3\sin^2\pi k_i.
$%\]
 This 3D system contains non-parallel potential defects and, hence,
 does not admit explicit procedure for finding
$\hat\Psi$ (see, e.g., \cite{Kjmaa1} for 2D case). There are various
methods to obtain the approximated solution of \er{e003}. We will use
the method based on the expansion of the operator \er{e003} in the basis $\cE^{\a}_{nm}$ to obtain the approximation of
$\hat\Psi$ with an arbitrary precision. Let $p\in\N$, $h=1/p$. Then, following \er{002} we have
\[\lb{e005}
 \O\ev[0,1)^3=\bigcup_{{\bf n}\in\mN}\{{\bf a}_{\bf n}+[0,h)^3\},
\ \
 \mN=\{0,..,p-1\}^3,\ \ {\bf a}_{\bf n}=h{\bf
 n}.
\]
It is convenient to use the indices ${\bf n},{\bf m}\in\mN$ in
\er{003} instead of numbers $i,j$. Introducing $A_{\bf n}=A(a_{\bf
n}+{\bf 1}h/2)$ (where ${\bf 1}=(1,1,1)$) we can write the
approximation of the equation $\cA\hat\Psi=1$ ($\cA$ denotes LHS of \er{e003}) as $\wt\cA\wt\Psi=1$, where
\[\lb{e006}
 \wt\cA=\sum_{{\bf n}\in\mN}\cE^{\es}_{{\bf n}{\bf n}}[A_{\bf n}]-
 \sum_{\substack{{\bf n},{\bf m}\in\mN\\n_2=m_2\\n_3=m_3}}
 \cE^{\{1\}}_{{\bf n}{\bf m}}[hV_1]-\sum_{\substack{{\bf n},{\bf m}\in\mN\\n_1=m_1\\n_3=m_3}}
 \cE^{\{2\}}_{{\bf n}{\bf m}}[hV_2]-\sum_{\substack{{\bf n},{\bf m}\in\mN\\n_1=m_1\\n_2=m_2}}
 \cE^{\{3\}}_{{\bf n}{\bf m}}[hV_3].
\]
Then the matrices ${\bf A}_{\a}\in\C^{p^3\ts p^3}$, see \er{006},
\er{007}, are defined by
\begin{multline}\lb{e007}
 {\bf A}_{\es}=\diag(A_{\bf n}),\ \ {\bf
 A}_{\{1\}}=-hV_1(\d_{n_2m_2}\d_{n_3m_3})_{{\bf n},{\bf m}\in\mN},\
 \ {\bf A}_{\{2\}}=-hV_2(\d_{n_1m_1}\d_{n_3m_3})_{{\bf n},{\bf
 m}\in\mN},\\
 {\bf A}_{\{3\}}=-hV_3(\d_{n_1m_1}\d_{n_2m_2})_{{\bf n},{\bf
 m}\in\mN},\ \ {\bf A}_{\{1,2\}}={\bf A}_{\{1,3\}}={\bf
 A}_{\{2,3\}}={\bf A}_{\{1,2,3\}}={\bf 0}.
\end{multline}
The matrices ${\bf B}_{\a}$, see \er{007}, are
\begin{multline}\lb{e008}
 {\bf B}_{\es}={\bf A}_{\es},\ \ {\bf B}_{\{i\}}={\bf A}_{\es}+{\bf A}_{\{i\}}\ (i=1,2,3),\\
 {\bf B}_{\{1,2\}}={\bf A}_{\es}+{\bf A}_{\{1\}}+{\bf A}_{\{2\}},
 \ \ {\bf B}_{\{1,3\}}={\bf A}_{\es}+{\bf A}_{\{1\}}+{\bf A}_{\{3\}},
\\
 {\bf B}_{\{2,3\}}={\bf A}_{\es}+{\bf A}_{\{2\}}+{\bf A}_{\{3\}},\ \ ,\ \ {\bf B}_{\{1,2,3\}}={\bf A}_{\es}+{\bf A}_{\{1\}}+{\bf
 A}_{\{2\}}+{\bf A}_{\{3\}}.
\end{multline}
To compute $\wt\cA^{-1}$ we follow \er{011} and \er{009}, \er{010}.
Define the matrices ${\bf C}_{\a}$ by
\begin{multline}\lb{e009}
 {\bf C}_{\es}={\bf B}_{\es}^{-1},\ \ {\bf C}_{\{i\}}={\bf B}_{\{i\}}^{-1}-{\bf B}_{\es}^{-1}\ (i=1,2,3),\ \
 {\bf C}_{\{1,2\}}={\bf B}_{\{1,2\}}^{-1}-{\bf B}_{\{1\}}^{-1}-{\bf B}_{\{2\}}^{-1}+{\bf
 B}_{\es}^{-1},\\
 {\bf C}_{\{1,3\}}={\bf B}_{\{1,3\}}^{-1}-{\bf B}_{\{1\}}^{-1}-{\bf B}_{\{3\}}^{-1}+{\bf
 B}_{\es}^{-1},\ \ {\bf C}_{\{2,3\}}={\bf B}_{\{2,3\}}^{-1}-{\bf B}_{\{2\}}^{-1}-{\bf B}_{\{3\}}^{-1}+{\bf
 B}_{\es}^{-1},
\\
 {\bf C}_{\{1,2,3\}}={\bf B}_{\{1,2,3\}}^{-1}-{\bf B}_{\{1,2\}}^{-1}-
 {\bf B}_{\{1,3\}}^{-1}-{\bf B}_{\{2,3\}}^{-1}+{\bf B}_{\{1\}}^{-1}+{\bf B}_{\{2\}}^{-1}+{\bf B}_{\{3\}}^{-1}-{\bf B}_{\es}^{-1}.
\end{multline}
All the matrices ${\bf B}_{\a}$ are invertible since $\wt\cA^{-1}$
exists (because $\wt\cA+i\ve$ is self-adjoint). Then $\wt\Psi$ has
the form
\begin{multline}\lb{e010}
\wt\Psi=\wt\cA^{-1}1=\pmb{\chi}^{\top}\lt({\bf
C}_{\es}\pmb{\chi}+p{\bf
C}_{\{1\}}\int_0^1\pmb{\chi}(x_1,k_2,k_3)dx_1+p{\bf
C}_{\{2\}}\int_0^1\pmb{\chi}(k_1,x_2,k_3)dx_2+\\
p{\bf C}_{\{3\}}\int_0^1\pmb{\chi}(k_1,k_2,x_3)dx_3+p^2{\bf
C}_{\{1,2\}}\int_0^1\int_0^1\pmb{\chi}(x_1,x_2,k_3)dx_1dx_2+ \\
p^2{\bf C}_{\{1,3\}}\int_0^1\int_0^1\pmb{\chi}(x_1,k_2,x_3)dx_1dx_3+
p^2{\bf
C}_{\{2,3\}}\int_0^1\int_0^1\pmb{\chi}(k_1,x_2,x_3)dx_2dx_3+\\
p^3{\bf
C}_{\{1,2,3\}}\int_0^1\int_0^1\int_0^1\pmb{\chi}(x_1,x_2,x_3)dx_1dx_2dx_3\rt),
\end{multline}
where the vector-valued function
$\pmb\chi=(\prod_{i=1}^3\chi_{n_i}(k_i))_{{\bf n}\in\cN}$ (see
also \er{014} for $\chi_n$). Due to the simple form of $\pmb{\chi}$, all the
integrals in \er{e010} can be computed explicitly: for example,
$$
 p^2\int_0^1\int_0^1\pmb{\chi}(x_1,k_2,x_3)dx_1dx_3=(\chi_{n_2}(k_2))_{{\bf
n}\in\cN}.
$$ 
So, $\wt\Psi$ in \er{e010} is given explicitly. The norm of
difference between $\cA$ \er{e003} and $\wt\cA$ \er{e006} depends on
how much $A_{\bf n}$ approximates $A$, at least
$\|\cA-\wt\cA\|_{L^2\to L^2}\le 24\pi h$, see also Theorem \ref{T2}. The norm of the inverse
operators $\|\cA^{-1}\|_{L^2\to L^2}\le\ve^{-1}$ and
$\|\wt\cA^{-1}\|_{L^2\to L^2}\le\ve^{-1}$ since $\cA+i\ve$ and
$\wt\cA+i\ve$ are self-adjoint. Hence
$\|\cA^{-1}-\wt\cA^{-1}\|_{L^2\to L^2}\le24\pi h/\ve^2$. In
particular,
\[\lb{e011}
 \|\hat\Psi-\wt\Psi\|_{L^2}\le 24\pi
h/\ve^2,\ \ \sum_{{\bf n}\in\Z^3}|\hat\Psi_{\bf
n}-\wt\Psi_{\bf n}|^2\le (24\pi h/\ve^2)^2,
\]
where $\wt\Psi_{\bf n}$ are Fourier coefficients of $\wt\Psi$. On
the side, we obtain the following spectral estimates
\[\lb{e012}
 \dist({\rm sp}(\cA),\bigcup_{\b\ss\{1,2,3\}}{\rm sp}({\bf B}_{\b}))\le24\pi
 h.
\]
Estimates \er{e011}, \er{e012} become better and better for $h\to0$. Depending on the number $|\a|$, the components ${\rm sp}({\bf
B}_{\a})\sm(\cup_{|\b|<|\a|}{\rm sp}({\bf B}_{\b}))$ approximate
the volume, planar, guided, and local isolated spectral components
of $\cA$ which correspond to bulk, planar, guided, and local wave
functions. These waves  propagate along the potential
defects of the corresponding dimension $3-|\a|$ and exponentially attenuate in the perpendicular directions to the defect, see, e.g., the corresponding discussion in \cite{Kjmaa1}.

\section{Proof of Theorems \ref{T1} and \ref{T2}}\lb{Proof}

At first, let us consider the semigroup of subsets
\[\lb{100}
 \mG=\{e_{\a}:\ \a\ss\{1,...,N\}\},\ \ e_{\a}e_{\b}=e_{\a\cup\b}
\]
and the corresponding $C^*$-algebra
\[\lb{101}
 \mM=\lt\{\sum_{\a\ss\{1,...,N\}}A_{\a}e_{\a}:\ A_{\a}\in\C\rt\}.
\]
The identity element in this algebra is $1=e_{\es}$, where $\es$ is
the empty set. All the basis elements ${\bf e}_{\a}^*={\bf e}_{\a}$ are self-adjoint. Define the mapping
\[\lb{102}
 \pi_1:\mM\to\C^{2^N},
\]
\[\lb{103}
 \pi_1\lt(\sum_{\a\ss\{1,...,N\}}A_{\a}e_{\a}\rt)=(B_{\a})_{\a\ss\{1,...,N\}},\
 \ B_{\a}=\sum_{\b\ss\a}A_{\b}.
\]
\begin{lemma}\lb{L1}
The mapping $\pi_1$ is the $C^*$-isomorphism. The inverse mapping
is defined by
\[\lb{104}
 \pi_1^{-1}((B_{\a})_{\a\ss\{1,...,N\}})=\sum_{\a\ss\{1,...,N\}}A_{\a}e_{\a},\
 \ A_{\a}=\sum_{\b\ss\a}(-1)^{|\a\sm\b|}B_{\b}.
\]
\end{lemma}
{\it Proof.} Consider the following basis in $\mM$
\[\lb{105}
 f_{\a}=e_{\a}\prod_{n\not\in\a}(1-e_{\{n\}}).
\]
Direct calculations give us
\[\lb{106}
 f_{\a}^*=f_{\a},\ \ \ f_{\a}f_{\b}=\d_{\a\b}f_{\a}.
\]
This means that $\{f_{\a}\}_{\a\ss\{1,...,N\}}$ can be considered as an orthogonal (in the algebraic sense) basis in $\C^{2^N}$ because the number of elements $\#\{\a:\ \a\ss\{1,...,N\}\}=2^N$. Hence, $\mM$ is isomorphic to $\C^{2^N}$ with the isomorphism $e_{\a}\leftrightarrow f_{\a}$. Using
\[\lb{107}
 \sum_{\a\ss\{1,...,N\}}A_{\a}e_{\a}=\sum_{\a\ss\{1,...,N\}}A_{\a}e_{\a}\prod_{n\not\in\a}(1-e_{\{n\}}+e_{\{n\}})
 =\sum_{\a\ss\{1,...,N\}}A_{\a}\sum_{\b\supset\a}e_{\b}\prod_{n\not\in\b}(1-e_{\{n\}})
\]
\[\lb{108}
 =\sum_{\a\ss\{1,...,N\}}A_{\a}\sum_{\b\supset\a}f_{\b}=
 \sum_{\a\ss\{1,...,N\}}(\sum_{\b\subset\a}A_{\b})f_{\a}=\sum_{\a\ss\{1,...,N\}}B_{\a}f_{\a}
\]
we obtain that the isomorphism $e_{\a}\leftrightarrow f_{\a}$
coincides with $\pi_1$ \er{102}-\er{103}. Similarly, identities
\[\lb{109}
 \sum_{\a\ss\{1,...,N\}}B_{\a}f_{\a}=\sum_{\a\ss\{1,...,N\}}B_{\a}e_{\a}\prod_{n\not\in\a}(1-e_{\{n\}})=
 \sum_{\a\ss\{1,...,N\}}B_{\a}\sum_{\b\supset\a}(-1)^{|\b\sm\a|}e_{\b}
\]
\[\lb{110}
 =\sum_{\a\ss\{1,...,N\}}(\sum_{\b\subset\a}(-1)^{|\a\sm\b|}B_{\b})e_{\a}=\sum_{\a\ss\{1,...,N\}}A_{\a}e_{\a}
\]
give us the form of the inverse mapping $\pi_1^{-1}$ \er{104}. \BBox

By Lemma \ref{L1}, the $C^*$-algebra $\mM_{M}=\mM^{M\ts M}$ of $M\ts M$
matrices with entries belonging to $\mM$ is isomorphic to $(\C^{M\ts
M})^{2^N}$. The corresponding isomorphism is
\[\lb{111}
 \pi_M:\mM_{M}\to(\C^{M\ts M})^{2^N},\ \ \pi_M(\sum_{\a\ss\{1,...,N\}}{\bf
 A}_{\a}e_{\a})=({\bf B}_{\a})_{\a\ss\{1,...,N\}},\ \ {\bf B}_{\a}=\sum_{\b\ss\a}{\bf
 A}_{\b},
\]
\[\lb{112}
 \pi_M^{-1}(({\bf B}_{\a})_{\a\ss\{1,...,N\}})=\sum_{\a\ss\{1,...,N\}}{\bf
 A}_{\a}e_{\a},\ \ {\bf A}_{\a}=\sum_{\b\ss\a}(-1)^{|\a\sm\b|}{\bf
 B}_{\b}.
\]
The same result holds for the $C^*$-algebra $\mM_{MS}\cong\mM_{M}^{S\ts
S}\cong\mM^{MS\ts MS}$. For any ${\bf A}\in\C^{M\ts M}$ and
$\a\ss\{1,...,N\}$, the basis elements $\cF_{ij}^{\a}[{\bf
A}]=(\d_{i\tilde{i}}\d_{j\tilde{j}}e_{\a}{\bf A})_{\tilde{i},\tilde{j}=1}^S\in\mM_{MS}$  satisfy the same
equations as the basis elements $\cE_{ij}^{\a}[{\bf A}]\in\cA$ (see
\er{005})
\[\lb{113}
 \cF_{ij}^{\a}[{\bf A}]^*=\cF_{ji}^{\a}[{\bf A}^*],\ \ \cF_{ij}^{\a}[{\bf A}]+\cF_{ij}^{\a}[{\bf B}]=\cF_{ij}^{\a}[{\bf A}+{\bf B}],\ \
 \cF_{ij}^{\a}[{\bf A}]\cF_{kl}^{\b}[{\bf B}]=\d_{jk}\cF_{il}^{\a\cup\b}[{\bf A}{\bf
 B}].
\]
This means that $\mA\cong\mM_{MS}$ with the natural isomorphism
$\cE_{ij}^{\a}[{\bf A}]\leftrightarrow\cF_{ij}^{\a}[{\bf A}]$. Then,
comparing \er{006}-\er{008}  and \er{111}-\er{112} (with $MS$
instead of $M$) we deduce that $\pi$ in \er{008} is the isomorphism satisfying
\er{009}-\er{010}.

Now, let us prove Theoem \ref{T2}. It is seen that
$$
 (\cA_c-\cA)u({\bf k})=\sum_{\a\ss\{1,...,N\}}\sum_{i=1}^{p^N}\int_{[0,1)^{|\a|}\cap\O_i}(A_{\a}({\bf k},{\bf x}_{\a})-A_{\a}(\wt{\bf k},(\wt{\bf x}_{i})_{\a}))u({\bf k}_{\ol{\a}}+{\bf x}_{\a})d{\bf x}_{\a},
$$
where $\wt{\bf k}$ is the center of the cube $\O_j$ for which ${\bf k}\in\O_j$, and $\wt {\bf x}_i$ is the center of $\O_i$. Then, the mean value estimates
$$
 \max_{{\bf x}\in\O_i}|A_{\a}({\bf k},{\bf x}_{\a})-A_{\a}(\wt{\bf k},(\wt{\bf x}_{i})_{\a})|\le\frac{N}{2p}\|\nabla A_{\a}\|_{L^{\iy}}
$$
finish the proof of \er{o18}. Estimates \er{o19} are classical spectral estimates for self-adjoint operator $\cA_c$ and self-adjoint perturbed operator $\cA=\cA_c+\cA-\cA_c$ in terms of the norm of perturbation $\d=\|\cA-\cA_c\|_{L^2\to L^2}$, see, e.g., \cite{Kato}. If $\cA_c$ is invertible then $\cA$ is invertible for all sufficiently large $p$ and $\|\cA^{-1}\|_{L^2\to L^2}\to\|\cA_c^{-1}\|_{L^2\to L^2}$ for $p\to\iy$. Moreover, using the Neumann series for the inverse of perturbed operator, we deduce that
$$
 \|\cA_c^{-1}-\cA^{-1}\|_{L^2\to L^2}=\|\lt(\sum_{n=1}^{\iy}(\cA^{-1}(\cA_c-\cA))^n\rt)\cA^{-1}\|_{L^2\to L^2}\le\frac{\d\|\cA^{-1}\|^2_{L^2\to L^2}}{1-\d\|\cA^{-1}\|_{L^2\to L^2}}.
$$ 
The last term tends to 0 since $\d$ tends to 0 for $p\to\iy$.

\section{Conclusion\lb{Conclusion}}
We have shown that the analysis of mixed multidimensional integral and some type of ergodic operators can be explicitly reduced to the analysis of special matrices. This allows us to compute functions of such operators and their spectra explicitly with an arbitrary precision.

%\section*{Ethics statement}
%This work did not involve any active collection of human data.
%
%\section*{Data accessibility statement}
%This work does not have any experimental data.
%
%\section*{Competing interests statement}
%I have no competing interests.
%
%\section*{Authors contributions}
%I am a single author. I proved all the results and wrote this work myself.

%\section*{Acknowledgements}
%This work was partially supported by the RSF project
%N\textsuperscript{\underline{o}}15-11-30007 and DFG project TRR 181.

\section*{Funding statement}
This paper is a contribution to the project M3 of the Collaborative Research Centre TRR 181 "Energy Transfer in Atmosphere and Ocean" funded by the Deutsche Forschungsgemeinschaft (DFG, German Research Foundation) - Projektnummer 274762653. This work is also supported by the RFBR (RFFI) grant No. 19-01-00094.

\bibliography{bibl_perp1}

\end{document}